\documentclass[sigconf]{acmart}
\renewcommand\footnotetextcopyrightpermission[1]{} % removes footnote with conference information in first column

\usepackage{booktabs} % For formal tables

\acmConference[MDE4IoT’18]{International Workshop on Model-Driven Engineering for the Internet-of-Things}{October 2018}{Copenhagen, Denmark}
\begin{document}
\title{ThingML+}
%\titlenote{Produces the permission block, and
%  copyright information}
\subtitle{Augmenting Model-Driven Software Engineering for the Internet of Things with Machine Learning}
%\subtitlenote{The full version of the author's guide is available as
%  \texttt{acmart.pdf} document}

\author{Armin Moin}
%\authornote{Dr.~Trovato insisted his name be first.}
\orcid{0000-0002-8484-7836}
\affiliation{%
  \institution{Technical University of Munich}
  \streetaddress{Boltzmannstr. 3}
  \city{Munich}
  \state{Germany}
  \postcode{85748}
}
\email{moin@in.tum.de}

\author{Stephan Rössler}
%\authornote{The secretary disavows any knowledge of this author's actions.}
\affiliation{%
  \institution{Software AG}
  \streetaddress{Landsberger Str. 155}
  \city{Munich}
  \state{Germany}
  \postcode{80687}
}
\email{stephan.roessler@softwareag.com}

\author{Stephan Günnemann}
%\authornote{This author is the
%  one who did all the really hard work.}
\affiliation{%
  \institution{Technical University of Munich}
\streetaddress{Boltzmannstr. 3}
\city{Munich}
\state{Germany}
\postcode{85748}
}
\email{guennemann@in.tum.de}

\renewcommand{\shortauthors}{A. Moin et al.}

\begin{abstract}
In this paper, we present the current position of the research project ML-Quadrat, which aims to extend the methodology, modeling language and tool support of ThingML - an open source modeling tool for IoT/CPS - to address Machine Learning needs for the IoT applications. Currently, ThingML offers a modeling language and tool support for modeling the components of the system, their communication interfaces as well as their behaviors. The latter is done through state machines. However, we argue that in many cases IoT/CPS services involve system components and physical processes, whose behaviors are not well understood in order to be modeled using state machines. Hence, quite often a data-driven approach that enables inference based on the observed data, e.g., using Machine Learning is preferred. To this aim, ML-Quadrat integrates the necessary Machine Learning concepts into ThingML both on the modeling level (syntax and semantics of the modeling language) and on the code generators level. We plan to support two target platforms for code generation regarding Stream Processing and Complex Event Processing, namely Apache SAMOA and Apama.
\end{abstract}

%	<ccs2012>
%	<concept>
%	<concept_id>10011007.10011006.10011066.10011070</concept_id>
%	<concept_desc>Software and its engineering~Application specific development environments</concept_desc>
%	<concept_significance>300</concept_significance>
%	</concept>
%	</ccs2012>
%\end{CCSXML}

%\ccsdesc[300]{Software and its engineering~Application specific development environments}


\maketitle

\section{Introduction}
\paragraph{The Internet of Things and the Role of Artificial Intelligence}
Currently, we are on the edge of the next industrial revolution, where the Internet of Things (IoT) and smart Cyber-Physical Systems (CPS) that are connected via the IoT, provide the infrastructure for that. The IoT is an expansion of the Internet into new domains, where constrained embedded devices such as sensors and actuators play an important role \cite{Atzori+2010}. Moreover, CPS are systems of systems, which have both physical and virtual, i.e., digital elements. \textit{Artificial Intelligence (AI)} is the key factor that distinguishes between the previous revolution in the industry world-wide, which led to industrial automation, and the current one leading to \textit{Cognitive Systems} that can possess \textbf{cognitive capabilities} such as \lq{}learning\rq{}.

\paragraph{Why Model-Driven Engineering for IoT/CPS?}
As mentioned above, smart Cyber-Physical Systems (CPS) are connected through the IoT. As these systems of systems are very large, highly distributed, very heterogeneous and cross-domain, there is an eminent need for \textbf{abstraction} and \textbf{automation} to be able to specify, design, develop, analyze, verify and maintain them in a cost-effective manner. One of the promising approaches that provides both abstraction and automation is the Model-Driven Engineering (MDE), also known as Model-Based Engineering, where models are the core elements in the entire life-cycle from the specification and design phase to the implementation, deployment and maintenance phases. \footnote{Although the border between these phases might not exist anymore in its classic sense for modern applications.} MDE has already proven quite successful in some domains such as Embedded Systems, e.g., in the automobile industry. Hence, it sounds like the most natural and the most suitable approach to address the said challenges in the domain of CPS/IoT. \cite{Schaetz2014}

\paragraph{ThingML: an open source state-of-the-art MDE solution for IoT/CPS}
When we talk about MDE, we actually mean Model-Driven Software Engineering (MDSE), particularly the Domain-Specific Modeling (DSM) approach to that. One such state-of-the-art solution for the CPS/IoT domain, which is available as free open source software, is ThingML \footnote{We chose ThingML due to the prior work of our industrial partner.}. ThingML is not only the name of the project, but also the name of the domain-specific modeling language, the methodology and the free open source tool supporting them. The (textual) model editor of ThingML supports the user (e.g., a CPS/IoT service developer) to model the distributed system using the following elements: (i) \textbf{components} (i.e., \textit{Things}) with asynchronous message passing interfaces (Ports), (ii) composite state machines aligned with the UML2 state charts for specifying the \textbf{behavior} of components, and (iii) an imperative \textbf{action language} for the event processing rules. This action language is platform-independent, but includes a template language for linking platform-specific models in an easy manner. Once the model is complete, code generators (also known as Model-to-Text or Model-to-Code transformations in the MDSE terminology) can be employed to automatically generate the full implementation for specific target platforms and communication protocols that are supported by ThingML. The generated implementation includes the source code and configuration scripts, and may also include documentation. Last but not least, the ThingML tool is built based on the free open source Eclipse Modeling Framework (EMF), thus highly extensible and interoperable. \cite{Harrand+2016,FleureyMorin2017}

\paragraph{Motivation \& Contribution of this Position Paper}
We argue that model-driven software engineering languages and tools for the IoT shall provide support for Machine Learning by design. Concretely, we propose a complementary view for specifying the behavior of components, i.e., \textit{things} in ThingML, which is not based on state machines in their current form, but based on \textit{inference} from the observed data. This means, we enable a data-driven approach for specifying the behavior. In other words, using this complementary view, one shall be able to model the Machine Learning algorithm at design-time and let the system partially or fully learn the behavior based on the observed data at run-time. In Section \ref{Position}, we illustrate our position. This comprises the comparison of models in Machine Learning and models in Software Engineering, our core idea on integrating them for the IoT applications, the advantages that we consider, and the challenges that we foresee for that. Finally, we conclude in Section \ref{conclusions}.

\section{Our Position} \label{Position}
\begin{figure}
	\centering
	\includegraphics[width=0.2\textwidth]{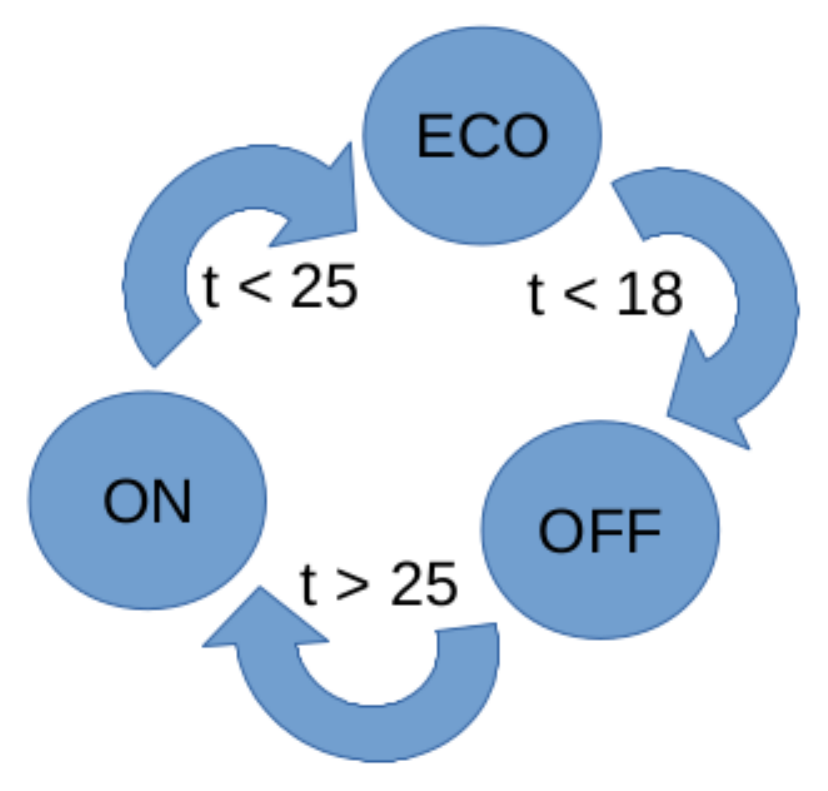}
	\caption{A sample state machine modeling the behavior of a smart Air Conditioner based on the room temperature.}
	\label{fig:AC}
\end{figure}

We believe that the Model-Driven Engineering (MDE) methodologies and tools for the Internet of Things (IoT) and smart Cyber-Physical Systems (CPS) must support the Artificial Intelligence (AI) needs of these applications both on the modeling level and also on the code generation level in an integrated and seamless manner. Hence, we propose an alternative view for modeling the behavior of components, i.e., things in ThingML. This new view enables users of the tool to delegate the definition of the behavior of the thing (i.e, system competent or IoT device) to the AI algorithms (specifically ML algorithms), which are able to conduct inference based on the observed data. In other words, for complex behaviors that are not easily understandable and specifiable via state machines, the ML algorithms can \textit{learn} the respective behaviors on their own in an effective and efficient manner. In addition, we provide an extension to the existing view for specifying behaviors using state machines so that advanced data analytics algorithms and methods can be employed for event detection and triggering state transitions. This latter contribution is inline with the contributions of the research project HEADS funded by the European Commission (FP7), where Complex Event Processing (CEP) capabilities have been introduced to the ThingML tool (also known as the HEADS IDE in the context of that project), e.g., using the CEP platform Apama. Figure \ref{fig:AC} depicts the graphical representation of a state machine that is currently used (with the existing ThingML tool) for modeling the behavior of a smart Air Conditioner using the data that comes through asynchronous message passing from a temperature sensor in the room.

\subsection{Models in Machine Learning vs. Software Engineering}
\paragraph{ML Models}
Machine Learning (ML) is currently used as a promising \textit{data-driven} approach in industry to address many complex problems. ML is one of the several fields that are widely used in data analytics. \footnote{It is perhaps the most important one. For instance, Deep Neural Networks, a family of ML algorithms are currently widely used in industry.} In data-driven approaches, we observe the data instances generated by some process, then we build some model (e.g., statistical) and \textit{train} that model via a ML algorithm using the data instances. Finally, we use the trained model in future, e.g., for making predictions about the possible outputs of that process. Hence, a model in ML is an abstraction / artifact that can help us in making inference based on the observed data, e.g., for making predictions. A popular example for this use case is predictive maintenance, where ML models can predict possible failures of the system in advance, e.g., based on anomaly detection. There exist various ways for categorizing ML models and algorithms. For instance, ML models fall from one point of view into two categories: parametric and non-parametric. In parametric models, we assume a specific functional form for the model (i.e., the statistical distribution that is assumed to be the generator of the observed data), where a small number of parameters control the form of the model. The linear regression model and the Neural Networks family are examples of the linear and nonlinear parametric models, respectively. However, in non-parametric models, the form of the model is defined by the size of the dataset. Although these models still contain parameters, their parameters do not affect the form of the model, but its complexity. One popular example of this category is the Support Vector Machines (SVM) family. Unlike parametric models, non-parametric ones usually keep part of the data instances or all of them for their future use. For this reason, they are also called memory-based or instance-based. Confusingly, some sources refer to the non-memory-based models as model-based. However, we do not use that term. In our view, all mentioned approaches for ML modeling are model-based. Last but not least, some diagrammatic representation of probability distributions used in ML, known as Probabilistic Graphical Models (PGM) has provided a very intuitive, sound and useful way of visualization and analysis of ML models. \cite{Bishop2006}

\paragraph{SE Models}
From the above explanation, it is clear that ML models are very much different than SE models. Popular examples of SE models can be found in the UML (Unified Modeling Language) standard. They can usually be categorized into structural models and behavioral models (including interactions). The models in ThingML are currently merely SE models. However, we plan to link them with ML models. In the following, we explain our position in more detail.

\subsection{Bringing ML Models and SE models Together}
\paragraph{It is NOT (only) about Model-based Machine Learning}
In his scientific article, \textit{Model-based Machine Learning} \cite{Bishop2013}, Christopher M. Bishop has already called for following a model-driven approach in the field of ML similar to the Model-Driven Software Engineering (MDSE) paradigm. Specifically, he has proposed the concept of \textit{probabilistic programming} and presented a Domain-Specific Modeling Language (DSML) called \textit{Infer.NET} for PGMs \cite{Bishop2013}.

\paragraph{It is NOT (only) about abstraction nor (merely) code generation}
There exist already various workflow designers and frameworks in the field of data analytics such as KNIME, RapidMiner, CamSaS Musketeer and TensorBoard that make conducting data analytics tasks more efficient. They provide a higher level of abstraction and support various target data analytics and data engineering platforms for partial or full code generation. Similarly, in the field of IoT, there exist several mashup tools that provide a higher level of abstraction for developing IoT services by combining data and services over the IoT. Note that many of these cloud-based tools go far beyond simple mashup and also offer other cloud services such as data analytics and AI capabilities. Examples include but are not limited to the startup waylay.io, Microsoft Azure and the IBM Watson IoT Platform. However, we propose a holistic approach that provides a methodology and tool support for systematic engineering of the entire software / smart services for the IoT/CPS applications. This includes also the data analytics (specifically ML) and IoT mashup capabilities. Our approach is based on the MDE paradigm. Note that this systematic approach is not the case in any of the said solutions. For instance, using the existing tools, one cannot separate the business logic and the underlying technologies and at the same time cover both the Data Science and Engineering as well as the Software Engineering aspects of the IoT/CPS applications.

\paragraph{It is NOT (necessarily) about Graphical / Visual Diagrams}
Note that when we talk about DSML and modeling in general, we do not necessarily mean graphical / visual diagrams. Model instances and model editors can be also textual. It is often a misunderstanding that modeling has necessarily something to do with graphical representations. However, we plan to make our DSML and modeling tool graphical / visual, in order to make it more intuitive and user-friendly.

\paragraph{Advantages}
Beside the typical advantages of MDE, and the specific advantages of the MDE for the domain of IoT/CPS, which by nature involves much more heterogeneity and much larger scale than classic application domains such as embedded systems (see, e.g., \cite{Harrand+2016,FleureyMorin2017}), we can offer the advantage of facilitating the employment of data analytics algorithms and methods on the modeling level and having the source code still automatically generated out of the model instances. This way, software engineers, who do not necessarily have deep knowledge and skills in the filed of Data Science and Engineering can easily create smart services for the IoT/CPS without mastering the algorithms (e.g., ML algorithms) and data analytics methods as well as the various underlying platforms (e.g., Spark, Storm, Flink, Samza). This will make an important contribution to the current problem of lack of Data Scientists in the industry world-wide. Note that the tool will provide some hints and advice during the modeling time to support the user of the tool in employing the ML algorithms and data analytics methods. Concretely, we plan to support the Apache SAMOA as well as the Apama platforms for code generation. However, our tool will be open source and fully extensible for further platforms and use cases.

\paragraph{Challenges}
Again, similar to the advantages, we do not mention the typical challenges of MDE and MDE for the IoT, but rather the specific ones for this work. Currently, we foresee the main challenge to be related to the very different natures of models in SE and ML. For instance, it sounds quite challenging to generate code out of the model instances that do not have state machines for behavior specification, but rather inference models. Moreover, ML models and algorithms themselves are also quite different and require very different measures, e.g, for data preparation. Further, often one cannot simply use the ML algorithms and methods just as black-boxes. Actually, one can do that, but there will be usually no good results. For instance, Neural Networks that are quite popular, are very sensitive to the initialization of their parameters and also to the architectural decisions including but not limited to the number of hidden layers, number of units per layer, and so forth. Since we do not yet have any formal specification for such practices, but instead they are conducted based on experiments by practitioners, it will be quite challenging to abstract such problems and tasks from the user and go for full automation. This also holds for data preparation and cleansing tasks. Such challenges are studied and addressed in a relatively new field of ML, known as AutoML.

\section{Conclusions}\label{conclusions}
In this paper, we presented the current position of the research project ML-Quadrat on the topic of MDE4IoT. We proposed supporting AI, particularly, Machine Learning, in modeling tools by design on the modeling level and on the code generators level. We believe that a holistic approach and a systematic methodology that covers both the SE and the ML aspects is needed. The project ML-Quadrat aims to realize this vision using the ThingML tool as the basis.

%\end{document}  % This is where a 'short' article might terminate

%\appendix
%Appendix A

\begin{acks}
This work is funded by the German Federal Ministry of Research and Education (BMBF) through the Software Campus initiative (project ML-Quadrat).
\end{acks}

\bibliographystyle{ACM-Reference-Format}
\bibliography{refs}

\end{document}